\documentclass[twocolumn,showpacs,aps,prl,superscriptaddress]{revtex4}

\usepackage{graphicx}
\usepackage{dcolumn}
\usepackage{epsfig}
\usepackage{psfrag}
\usepackage{amsmath}

\newcommand{\BaBarYear}    {10}
\newcommand{\BaBarNumber}  {003}

\newcommand{\SLACPubNumber} {14026}
\newcommand{\LANLNumber} {1004.0240 [hep-ex]}
   
 \newcommand{\BaBarType}      {PUB}  

\input pubboard/babarsym   

\RequirePackage{xspace}

\newcommand{\calP}{\ensuremath{{\cal P}}}

\newcommand{\pvec}{{\bf p}}

\newcommand{\acp}{\ensuremath{\calA_{\rm ch}}}

\newcommand{\calB}{\ensuremath{{\cal B}}}

\newcommand{\DE}{\ensuremath{\Delta E}}

\newcommand{\xf}{\ensuremath{{\cal F}}}

\newcommand\etal{{\it et al.}}

\newcommand{\msp}{\ensuremath{\phantom{-}}}

\newcommand{\bfig}{\begin{figure}[htbpc!]}
\newcommand{\efig}{\end{figure}}
\newcommand\bef{\begin{figure}}
\newcommand\edf{\end{figure}}
\newcommand\dbline{\noalign{\vskip 0.10truecm\hrule}\noalign{\vskip 2pt}\noalign{\hrule\vskip 0.10truecm}}

\newcommand\sgline{\noalign{\vskip 0.10truecm\hrule\vskip 0.10truecm}}
\newcommand\beq{\begin{equation}}
\newcommand\eeq{\end{equation}}
\newcommand\bear{\begin{array}}
\newcommand\enar{\end{array}}
\newcommand\beqa{\begin{eqnarray}}
\newcommand\eeqa{\end{eqnarray}}
\newcommand\ben{\begin{enumerate}}
\newcommand\een{\end{enumerate}}

\newcommand{\etatogg}{\ensuremath{\eta\ra\gaga}}

\newcommand{\etapepp}{\ensuremath{\etapr_{\eta\pi\pi}}}
\newcommand{\etaptoepp}{\ensuremath{\etapr\ra\eta\pip\pim}}

\newcommand{\etaprg}{\ensuremath{\etapr_{\rho\gamma}}}
\newcommand{\etaptorg}{\ensuremath{\etapr\ra\rho^0\gamma}}

\newcommand{\Kst}{\ensuremath{K^*}}
\newcommand{\Kzerst}{\ensuremath{\Kstar_0(1430)}}
\newcommand{\Ktwost}{\ensuremath{\Kstar_2(1430)}}
\newcommand{\Kstp}{\ensuremath{\Kstarp}}
\newcommand{\Kzerstp}{\ensuremath{\Kstar_0(1430)^+}}
\newcommand{\Ktwostp}{\ensuremath{\Kstar_2(1430)^+}}
\newcommand{\Kstz}{\ensuremath{\Kstarz}}
\newcommand{\Kzerstz}{\ensuremath{\Kstar_0(1430)^0}}
\newcommand{\Ktwostz}{\ensuremath{\Kstar_2(1430)^0}}
   \newcommand{\KstpKppiz}{\ensuremath{\Kstarp_{K^+\pi^0}}}
   \newcommand{\KstptoKppiz}{\ensuremath{\Kstarp\ra K^+\pi^0}}
   \newcommand{\KstpKspip}{\ensuremath{\Kstarp_{\KS\pi^+}}}
   \newcommand{\KstptoKspip}{\ensuremath{\Kstarp\ra \KS\pi^+}}
   
   \newcommand{\KstztoKppim}{\ensuremath{\Kstarz\ra K^+\pi^-}}

   \newcommand{\rhop}{\ensuremath{\rho^+}}
   \newcommand{\rhopipiz}{\ensuremath{\rhop\ra \pi^+\pi^0}}

   \newcommand{\rhoz}{\ensuremath{\rho^0}}
   \newcommand{\rhozpipi}{\ensuremath{\rhoz\ra \pi^+\pi^-}}

\newcommand{\kstopipi}{\ensuremath{\KS\to\pi^+\pi^-}}

  \newcommand{\KstOne}{\ensuremath{K^{*}(892)}}

\newcommand{\fetapKst}{\ensuremath{\etapr K^{*}}\xspace}

\newcommand{\fetapKzst}{\ensuremath{\etapr\Kzerst}}
\newcommand{\etapKzst}{\ensuremath{\B\ra\fetapKzst}}
\newcommand{\fetapKtst}{\ensuremath{\etapr\Ktwost}}
\newcommand{\etapKtst}{\ensuremath{\B\ra\fetapKtst}}

\newcommand{\fetapKstp}{\ensuremath{\etapr\Kstp}}

\providecommand{\fetapKzstp}{\ensuremath{\etapr\Kzerstp}}
\providecommand{\etapKzstp}{\ensuremath{\Bp\ra\fetapKzstp}}
\newcommand{\BetapKzstp}{\ensuremath{\calB(\etapKzstp)}}
\providecommand{\fetapKtstp}{\ensuremath{\etapr\Ktwostp}}

\newcommand{\retapKstp}{\ensuremath{xx^{+xx}_{-xx}\pm xx}}

\newcommand{\uletapKstp}{\ensuremath{xx}}

\newcommand{\setapKstp}{\ensuremath{xx}}

   \newcommand{\fetapeppKstpKppiz}{\ensuremath{\etapr_{\eta\pi\pi}\Kstarp_{\Kp\piz}}}
   
   \newcommand{\fetaprgKstpKppiz}{\ensuremath{\etapr_{\rho\gamma}\Kstarp_{\Kp\piz}}}
   \newcommand{\etaprgKstpKppiz}{\ensuremath{\Bp\ra\fetaprgKstpKppiz}}

\newcommand{\fetapeppKtstpKspip}{\ensuremath{\etapr_{\eta\pi\pi}\Ktwostp_{\KS\pip}}}

\newcommand{\fetapeppKtstpKppiz}{\ensuremath{\etapr_{\eta\pi\pi}\Ktwostp_{\Kp\piz}}}

\newcommand{\fetaprgKtstpKspip}{\ensuremath{\etapr_{\rho\gamma}\Ktwostp_{\KS\pip}}}

\newcommand{\fetaprgKtstpKppiz}{\ensuremath{\etapr_{\rho\gamma}\Ktwostp_{\Kp\piz}}}

\newcommand{\fetapKstz}{\ensuremath{\etapr\Kstz}}

\providecommand{\fetapKzstz}{\ensuremath{\etapr\Kzerstz}}
\providecommand{\etapKzstz}{\ensuremath{\Bz\ra\fetapKzstz}}
\newcommand{\BetapKzstz}{\ensuremath{\calB(\etapKzstz)}}
\providecommand{\fetapKtstz}{\ensuremath{\etapr\Ktwostz}}

\newcommand{\retapKstz}{\ensuremath{xx^{+xx}_{-xx}\pm xx}}

\newcommand{\uletapKstz}{\ensuremath{xx}}

\newcommand{\setapKstz}{\ensuremath{xx}}

\providecommand{\fetapKst}{\ensuremath{\etapr\Kstar}}
   \newcommand{\fetapeppKstz}{\ensuremath{\etapr_{\eta\pi\pi}\Kstz}}
   
   \newcommand{\fetaprgKstz}{\ensuremath{\etapr_{\rho\gamma}\Kstz}}

   \newcommand{\fetapeppKtstz}{\ensuremath{\etapr_{\eta\pi\pi}\Ktwostz}}
   
   \newcommand{\fetaprgKtstz}{\ensuremath{\etapr_{\rho\gamma}\Ktwostz}}

\newcommand{\fetaprho}{\ensuremath{\etapr\rho}\xspace}

\newcommand{\fetaprhop}{\ensuremath{\etapr\rho^+}}
\newcommand{\etaprhop}{\ensuremath{\Bp\ra\fetaprhop}}

\newcommand{\retaprhop}{\ensuremath{xx^{+xx}_{-xx}\pm xx}}

\newcommand{\Aetaprhop}{\ensuremath{xx\pm xx \pm xx}}
\newcommand{\setaprhop}{\ensuremath{xx}}

\newcommand{\fetaprhoz}{\ensuremath{\etapr\rho^0}}
\newcommand{\etaprhoz}{\ensuremath{\Bz\ra\fetaprhoz}}

\newcommand{\retaprhoz}{\ensuremath{xx^{+xx}_{-xx}\pm xx}}

\newcommand{\uletaprhoz}{\ensuremath{xx}}

\newcommand{\setaprhoz}{\ensuremath{xx}}

\newcommand{\fetapfz}{\ensuremath{\etapr f_0}\xspace}
\newcommand{\etapfz}{\ensuremath{\Bz\ra\fetapfz}\xspace}
\newcommand{\Betapfz}{\ensuremath{\calB(\etapfz)}\xspace}
\newcommand{\retapfz}{\ensuremath{xx^{+xx}_{-xx}\pm xx}\xspace}

\newcommand{\uletapfz}{\ensuremath{xx}\xspace}

\newcommand{\setapfz}{\ensuremath{xx}\xspace}

\newcommand{\fomegaKst}{\ensuremath{\omega\Kstar}}
\newcommand{\omegaKst}{\ensuremath{\B\ra\fomegaKst}}

\providecommand{\fomegaKst}{\ensuremath{\omega\Kstar}}

\newcommand{\fzero}{\ensuremath{f_0}}

\newcommand{\KpiSwave}{\ensuremath{(K\pi)^*_0}}
\newcommand{\KpiSwavez}{\ensuremath{(K\pi)^{*0}_0}}
\newcommand{\KpiSwavep}{\ensuremath{(K\pi)^{*+}_0}}
\newcommand{\fetapKpiSwavez}{\ensuremath{\etapr\KpiSwavez}}

\newcommand{\fetapKpiSwavep}{\ensuremath{\etapr\KpiSwavep}}

\newcommand{\fetapeppKpiSwavez}{\ensuremath{\etapepp\KpiSwavez}}

\newcommand{\fetapeppKpiSwavepKppiz}{\ensuremath{\etapepp(\Kp\piz)^{*+}_0}}
\newcommand{\fetapeppKpiSwavepKspip}{\ensuremath{\etapepp(\KS\pip)^{*+}_0}}

\newcommand{\fetaprgKpiSwavez}{\ensuremath{\etaprg\KpiSwavez}}

\newcommand{\fetaprgKpiSwavepKppiz}{\ensuremath{\etaprg(\Kp\piz)^{*+}_0}}
\newcommand{\fetaprgKpiSwavepKspip}{\ensuremath{\etaprg(\KS\pip)^{*+}_0}}

\newcommand{\fetapKstV}{\ensuremath{\etapr\Kst(892)}}

\newcommand{\fetapKstzV}{\ensuremath{\etapr\Kst(892)^0}}
\newcommand{\fetapeppKstzV}{\ensuremath{\etapr_{\eta\pi\pi}\Kst(892)^0}}
\newcommand{\fetaprgKstzV}{\ensuremath{\etapr_{\rho\gamma}\Kst(892)^0}}

\newcommand{\fetapKstpV}{\ensuremath{\etapr\Kst(892)^+}}
\newcommand{\fetapeppKstpVKppiz}{\ensuremath{\etapr_{\eta\pi\pi}\Kst(892)^+_{\Kp\piz}}}
\newcommand{\fetaprgKstpVKppiz}{\ensuremath{\etapr_{\rho\gamma}\Kst(892)^+_{\Kp\piz}}}
\newcommand{\fetapeppKstpVKspip}{\ensuremath{\etapr_{\eta\pi\pi}\Kst(892)^+_{\KS\pip}}}
\newcommand{\fetaprgKstpVKspip}{\ensuremath{\etapr_{\rho\gamma}\Kst(892)^+_{\KS\pip}}}

\renewcommand{\retaprhop}{\ensuremath{9.7^{+1.9}_{-1.8}\pm 1.1}} 
\renewcommand{\setaprhop}{\ensuremath{5.8}}             
\renewcommand{\Aetaprhop}{\ensuremath{\msp0.26\pm 0.17\pm 0.02}}      

\renewcommand{\retaprhoz}{\ensuremath{1.5\pm 0.8\pm 0.3}} 
\renewcommand{\setaprhoz}{\ensuremath{2.0}}     
\renewcommand{\uletaprhoz}{\ensuremath{2.8}}    

\renewcommand{\retapfz}{\ensuremath{0.2^{+0.4}_{-0.3}\pm 0.1}} 
\renewcommand{\setapfz}{\ensuremath{0.5}}       
\renewcommand{\uletapfz}{\ensuremath{0.9}}      

\renewcommand{\retapKstz}{\ensuremath{3.1^{+0.9}_{-0.8}\pm 0.3}} 
\renewcommand{\setapKstz}{\ensuremath{4.0}}             
\renewcommand{\uletapKstz}{\ensuremath{4.4}}    
\newcommand{\AetapKstz}{\ensuremath{\msp0.02\pm 0.23\pm 0.02}}        

\newcommand{\retapKpiSwavez}{\ensuremath{7.4^{+1.5}_{-1.4}\pm 0.6}}
\newcommand{\retapKzstz}{\ensuremath{6.3 \pm 1.3 \pm 0.5 \pm 0.7}}
\providecommand{\setapKpiSwavez}{\ensuremath{5.6}}              
\newcommand{\RetapKzstz}{\ensuremath{(\retapKzstz)\times 10^{-6}}}

\newcommand{\AetapKpiSwavez}{\ensuremath{-0.19\pm 0.17\pm 0.02}}

\newcommand{\retapKtstz}{\ensuremath{13.7^{+3.0}_{-2.9}\pm 1.2}}
\providecommand{\setapKtstz}{\ensuremath{5.3}}          

\newcommand{\AetapKtstz}{\ensuremath{\msp0.14\pm 0.18\pm 0.02}}

\renewcommand{\retapKstp}{\ensuremath{4.8^{+1.6}_{-1.4}\pm 0.8}} 
\renewcommand{\setapKstp}{\ensuremath{3.8}}             
\renewcommand{\uletapKstp}{\ensuremath{7.2}}

\newcommand{\AetapKstp}{\ensuremath{-0.26\pm 0.27\pm 0.02}}       

\newcommand{\retapKpiSwavep}{\ensuremath{6.0^{+2.2}_{-2.0} \pm 0.9}} 
\providecommand{\setapKpiSwavep}{\ensuremath{2.9}}              
\newcommand{\retapKzstp}{\ensuremath{5.2 \pm 1.9 \pm 0.8 \pm 0.6}}
\newcommand{\RetapKzstp}{\ensuremath{(\retapKzstp)\times 10^{-6}}}

\newcommand{\uletapKpiSwavep}{\ensuremath{9.3}}

\newcommand{\AetapKpiSwavep}{\ensuremath{\msp0.06\pm 0.20\pm 0.02}}

\newcommand{\retapKtstp}{\ensuremath{28.0^{+4.6}_{-4.3}\pm 2.6}} 
\providecommand{\setapKtstp}{\ensuremath{7.2}}          

\newcommand{\AetapKtstp}{\ensuremath{\msp0.15\pm 0.13\pm 0.02}}

\newcommand{\theTitle}{{\boldmath \B-meson decays to $\etapr\rho$, 
$\etapr\fzero$, and $\etapr\Kst$}}

\begin{document}

\begin{flushleft}
\babar-\BaBarType-\BaBarYear/\BaBarNumber \\
SLAC-PUB-\SLACPubNumber \\
arXiv:\LANLNumber
\end{flushleft}

\title{\theTitle}

%
\author{P.~del~Amo~Sanchez}
\author{J.~P.~Lees}
\author{V.~Poireau}
\author{E.~Prencipe}
\author{V.~Tisserand}
\affiliation{Laboratoire d'Annecy-le-Vieux de Physique des Particules (LAPP), Universit\'e de Savoie, CNRS/IN2P3,  F-74941 Annecy-Le-Vieux, France}
\author{J.~Garra~Tico}
\author{E.~Grauges}
\affiliation{Universitat de Barcelona, Facultat de Fisica, Departament ECM, E-08028 Barcelona, Spain }
\author{M.~Martinelli$^{ab}$}
\author{A.~Palano$^{ab}$ }
\author{M.~Pappagallo$^{ab}$ }
\affiliation{INFN Sezione di Bari$^{a}$; Dipartimento di Fisica, Universit\`a di Bari$^{b}$, I-70126 Bari, Italy }
% ...
\collaboration{The \babar\ Collaboration}
\noaffiliation

\date{\today}

\begin{abstract}
We present measurements of \B-meson decays to the final states
\fetaprho, \fetapfz, and \fetapKst, where \Kst\ stands for a
vector, scalar, or tensor strange meson. 
We observe a significant signal or evidence for \fetaprhop\ and 
all the \fetapKst\ channels. We also measure, 
where applicable, the charge asymmetries, finding results
consistent with no direct \CP\-violation in all cases.
The measurements are performed on a data sample consisting of
$467 \times 10^6$ \BB\ pairs, collected with the \babar\ detector
at the PEP-II \epem\ collider at the SLAC National Accelerator
Laboratory. 
Our results favor the theoretical predictions from perturbative
QCD and QCD Factorization and we observe an enhancement of the
tensor \Ktwost\ with respect to the vector \KstOne\ component.
\end{abstract}


\maketitle

Charmless two-body decays of \B\ mesons are a powerful probe for
testing the standard model (SM) and searching for new physics
phenomena~\cite{cheng_smith}. Decays to final states containing $\eta$ or \etapr\
mesons exhibit a distinctive pattern of interference among the
dominant amplitudes and are also sensitive to a potentially large
flavor-singlet contribution~\cite{beneke_fs}. \B\-meson decays to 
the final states \fetaprho\ and \fetapKstV\ have been investigated 
theoretically within the SM
by means of perturbative QCD (pQCD) \cite{liu_pQCD}, QCD Factorization 
(QCDF) \cite{beneke_qcdf}, Soft Collinear Effective Theory (SCET)
\cite{wang_scet}, and $SU(3)$ flavor symmetry \cite{chiang_su3}. The predicted 
branching fractions to the final states $\etapr\Kst(892)^+$ and $\etapr\Kst(892)^0$ 
are in the range of a few times $10^{-6}$, whereas the branching fraction for
\etaprhoz\ is suppressed to $10^{-7}-10^{-8}$. There is some 
disagreement on the predictions of the branching fraction for
\etaprhop: SCET calculations give a value of $0.4 \times
10^{-6}$, whereas pQCD and QCDF predict a branching fraction of
$(6-9) \times 10^{-6}$.
There are no theoretical predictions for the branching fraction of \etapfz.

Experimentally, searches for these decays have been performed by the
\babar\ \cite{babar_epKst} and Belle \cite{belle_epKst} collaborations.
The former claimed evidence for the \fetaprhop, $\etapr\Kst(892)^+$, and 
$\etapr\Kst(892)^0$ final states, while the latter establishes upper limits that
are in poor agreement with the branching fractions, in the range 
$(4 - 9) \times 10^{-6}$, measured by \babar.

Very few predictions exist for $B$-meson decays to \fetapKzst\ and
\fetapKtst. In Ref.~\cite{liu_Kzst}, based on pQCD, the branching
fraction of \etapKzst\ is predicted to be in the range
$(20-80) \times 10^{-6}$, while in Ref.~\cite{munoz_Ktst} the branching 
fraction of \etapKtst\ is calculated to be $\sim 2 \times 10^{-6}$,
exploiting QCDF calculations.
No searches for the \fetapKzst\ and \fetapKtst\ final states 
have been reported. In a recent study of \omegaKst\ decays
\cite{babar_omegaKst}, where \Kst\ denotes the spin 0, 1, or 2 
states \Kzerst, \Kst (892), and \Ktwost, the tensor and scalar 
\Kst\ components were found to be significantly larger than the 
vector \Kst(892), a fact which was not anticipated by theory.

In this paper, we report measurements of the branching fractions of 
\B\ mesons decaying to the final states \fetaprho, \fetapfz, and \fetapKst.
Where applicable, we also measure the charge asymmetry
$\acp \equiv (\Gamma^- - \Gamma^+)/(\Gamma^- + \Gamma^+)$, where
the superscript to the decay width $\Gamma$ refers to the charge of 
the \Bpm\ meson or to the charge of the kaon in the neutral \B\ decays.

For this analysis we use the full \babar\ dataset, collected at the
PEP-II asymmetric-energy \epem\ collider located at the SLAC National
Accelerator Laboratory, consisting of $467 \times 10^6$ \BB\ pairs
originating from the decay of the \FourS\ resonance, produced at a
center-of-mass (CM) energy $\sqrt{s} = 10.58$ GeV. The \babar\
detector is described in detail elsewhere \cite{BABARNIM}.

We reconstruct $B$-daughter candidates through their decays
\etaptoepp\ (\etapepp), \etaptorg\ (\etaprg), \etatogg, \rhopipiz, 
\rhozpipi, $\fzero(980)\ra\pi^+\pi^-$ \cite{fzero}, \KstztoKppim,
\KstptoKppiz (\KstpKppiz), \KstptoKspip (\KstpKspip),
\kstopipi, and $\piz\ra\gaga$. We use both \etaptoepp\ and \etaptorg\ 
for \fetapKst, but due to the larger backgrounds affecting 
the \etaprg\ channel, we do not use the \etaptorg\ decay mode for the 
\fetaprho\ and \fetapfz\ final states. We require the invariant
masses of the $B$-daughter candidates to satisfy the
following requirements: $910 < m(\etapr) < 1000$ MeV,
$510 < m(\eta) < 580$ MeV, $510 < m(\rho/\fzero) < 1060$ MeV,
$750 < m(\Kst) < 1550$ MeV, $488 < m(\KS) < 508$ MeV, and
$120 < m(\piz) < 150$ MeV; these mass intervals are chosen 
to include sidebands to estimate the background levels. We require 
the photon energy of the \piz\ candidates to be greater than 
30 MeV in the laboratory frame, while the minimum energy is
100 MeV for the photons of $\eta$ candidates and 200 MeV for
the photons from the \etaptorg\ decay. A 
fit constraining the two pion tracks from the \KS\ decay 
to originate from the same decay vertex is performed and a
\KS\ candidate is selected if its flight length exceeds
three times its uncertainty. Daughter particles from \etapr,
$\eta$, $\rho$, \fzero, and \Kst\ decays are rejected if their
particle identification signatures are consistent with
those of electrons or protons; $K^+$ candidates are required
to be positively identified as kaons, and pions must fail
this criterion. Unless otherwise stated, charge-conjugate
reactions are always implied.

A $B$-meson candidate is characterized by two kinematic
variables, whose small correlation is accounted for in the
correction of the fit bias: 
the energy substituted mass $\mes\equiv\sqrt{s/4-\pvec_B^2}$
and the energy difference $\DE \equiv E_B-\sqrt{s}/2$, where 
$(E_B,\pvec_B)$ is the \B-meson four-momentum vector 
in the \FourS\ rest frame. Signal events peak at 0 in \DE\ and
at the \B\ mass \cite{PDG2008} in \mes, with a resolution in
\DE\ (\mes) of 20-35 MeV (3 MeV). We select events with
$5.25 < \mes < 5.29$ GeV and $|\DE| < 0.2$ GeV. 

The dominant backgrounds arise from random combinations of
particles in continuum $\epem\ra\qqbar$ 
events ($q=u,d,s,c$). The angle $\theta_T$ between the thrust 
axis \cite{thrust} of the $B$ candidate in the \FourS\
rest frame and that of the remaining particles in the event
is used to suppress this background. Jet-like continuum events
peak at $|\cos \theta_T|$ close to 1, while spherical \B\ decays
exhibit a flat distribution for this variable. Further rejection 
is achieved by restricting the range of the helicity angle of 
the $\rho$ or \Kst\ meson.
The helicity angle $\theta_{\cal H}$ is defined in the rest frame 
of the $\rho$ or \Kst\ and corresponds to the angle between two 
vectors: one is the momentum of the daughter pion and the other 
is the direction of the boost into this rest frame from the $B$ meson 
rest frame; for the \rhop\ candidate we use the positively charged pion. 
Table \ref{tab:cuts} summarizes the requirements we apply on $|\cos \theta_T|$ 
and $\cos \theta_{\cal H}$. After the selection criteria have
been applied, the average number of combinations per 
event in data is between 1.1 and 1.3, and we select the candidate 
with the highest $\chi^2$ probability in a geometric fit to 
a common \B-decay vertex. 
In this way the probability of selecting the correctly reconstructed 
event is a few percent higher with respect to a random selection.

\begin{table}[btp]
\begin{center}
\caption{
Selection requirements on $\cos \theta_T$ and $\cos \theta_{\cal H}$
for the different modes.
}
\label{tab:cuts}
\begin{tabular}{lccc}
\dbline
State               &  $|\cos \theta_T|$   & &  $\cos \theta_{\cal H}$ range \\
\dbline                                        
\fetaprhoz/\fzero   & $< 0.9$ & & $[-0.95,0.95]$ \\
\fetaprhop\         & $< 0.9$ & & $[-0.80,1]$ \\
\sgline
\fetapeppKstz\      & $< 0.9$ & & $[-0.85,1]$ \\
\fetaprgKstz\       & $< 0.6$ & & $[-0.85,1]$ \\
\fetapeppKstpKppiz\ & $< 0.9$ & & $[-0.75,1]$ \\
\fetaprgKstpKppiz\  & $< 0.7$ & & $[-0.75,1]$ \\
\fetapeppKstpKppiz\ & $< 0.9$ & & $[-0.90,1]$ \\
\fetaprgKstpKppiz\  & $< 0.7$ & & $[-0.90,1]$ \\
\dbline
\end{tabular}
\vspace{-5mm}
\end{center}
\end{table}

Further signal/background separation is provided by a Fisher discriminant
\xf\ exploiting four variables sensitive to the production dynamics and event 
shape: the polar angles, with respect to the beam axis in the \epem\ CM 
frame, of the $B$ candidate momentum and of the $B$ thrust axis; 
and the zeroth and second angular moments $L_{0,2}$ of the energy 
flow, excluding the $B$ candidate. 
The moments are defined by $L_j = \sum_i p_i\times\left|\cos\theta_i\right|^j,$ 
where $i$ labels a track or neutral cluster, $\theta_i$ is its angle
with respect to the $B$ thrust axis, and $p_i$ is its momentum.
The mean of \xf\ is shifted so that it is independent of the $B$-flavor 
tagging category \cite{ccbarK0} for \qqbar\ background. The Fisher variable 
provides about one standard deviation discrimination between $B$-decay events 
and continuum background.

We obtain the yields and the charge asymmetries \acp\ from extended
maximum-likelihood (ML) fits to the six observables: \DE, \mes, \xf,
the masses of the two resonance candidates 
$m_{\etapr}$ and $m_{\rho/\fzero/\Kst}$, and $\cos \theta_{\cal H}$. 
The fits distinguish among several categories: \qqbar\ background, 
\BB\ background, and signal component(s) (one for \fetaprhop, two 
for \fetaprhoz/\fzero, and three for the \fetapKst\ modes). For each 
event $i$ and category $j$ we define the probability density functions 
(PDFs) $\calP_j(x)$ for the variable $x$, with the resulting likelihood 
$\cal L$:

\begin{eqnarray}
\calP^i_j  &=& \calP_j(\mes^i) \calP_j(\DE^i) \calP_j(\xf^i) \label{eq:pdf} \\ 
&&\calP_j(m^i_{\etapr}) \calP_j(m^i_{\rho/\fzero/\Kst}) \calP_j(\cos \theta^i_{\cal H})\, , \nonumber \\
{\cal L} &=& \frac{e^{-\sum_j Y_j}}{N!} \prod_{i=1}^N \sum_j Y_j \calP_j^i\ , \label{eq:totalL}
\end{eqnarray}
where $Y_j$ is the yield for category $j$ and $N$ is the number of
events entering the fit. Where applicable, we split the yields
by the flavor of the decaying $B$ meson in order to measure \acp.
We find correlations among the observables to be significant in the
\BB\ components, whereas those are small in the data
samples, which contain mostly \qqbar\ background. An exception occurs 
in the \etaprgKstpKppiz\ mode, for which the correlation between 
$m_{\Kst}$ and $\cos \theta_{\cal H}$, if not taken into account, 
may cause large biases in the yield of the different \Kst\ components. 
In this case the factors for $m^i_{\Kst}$ and $\cos \theta^i_{\cal H}$ 
in Eq.~\ref{eq:pdf} are replaced with a two-dimensional non-parametric 
PDF \cite{keys}. We discuss below our treatment of the somewhat larger
correlations found in simulated signal events.

The signal component is studied from the Monte Carlo (MC) simulation 
\cite{geant} of the decay process and the response of the detector and
reconstruction chain. Signal events selected in simulation contain 
both properly reconstructed and incorrectly reconstructed $B$-meson 
candidates, which are labeled as "self-crossfeed" (SXF). SXF occurs 
either when some particles from the correct parent $B$ meson are 
incorrectly assigned to intermediate resonances or when particles from 
the rest of the event are used in the signal $B$ reconstruction.
The fraction of SXF events ranges between 14\% and 32\% and
we do not separate correctly reconstructed events from SXF in the fit. 
For the scalar $K\pi$ component, we use the LASS model 
~\cite{LASS,Latham}, which consists of the $\Kst_0(1430)$ resonance together 
with an effective-range nonresonant component. Backgrounds arising from 
\BB\ decays to charmless final states are modeled from the simulation.
We select the channels (20-40 for each final state) which have a high 
probability of passing our selection and build a sample of simulated specimens 
of these, weighting each component by its branching fraction, either 
measured or estimated. We model from this sample the PDFs for the \BB\ 
background component and fix its yield (25-350 events, depending on the final 
state) to the MC prediction. Backgrounds arising from $b\ra c$ transitions
have distributions very similar to those of \qqbar\ events and thus
they are absorbed by the continuum component. For the \fetaprho\ modes, 
we estimate the contribution of the nonresonant $\etapr \pi \pi$ with a fit 
of the data, selecting only the central region (which excludes the region
across the $\rho$ and \fzero\ resonances) of the $\etapr \pi \pi$ 
Dalitz plot. The expected nonresonant component is then included
in the charmless \BB\ background PDF. This procedure is not necessary 
for the \fetapKst\ fits, since the nonresonant $\etapr K \pi$ component 
is already included in the LASS model.

PDF shapes for the signals and \BB\ background are determined from
fits to MC samples, while for the \qqbar\ component we use data
sidebands, which we obtain excluding the signal region $5.27< \mes < 5.29$ GeV and
$|\DE| < 0.075$ GeV. The calibration of \mes\ and \DE\ is checked
by means of large data control samples of \B\ decays to 
charmed mesons with topology similar to the decays under study 
(e.g. $B^+\ra \Dzb(K^+\pim\piz)\pip$, $B^+\ra \Dzb(K^+\pim\piz)\rhop$).

We use a combination of Gaussian, exponential, and polynomial
functions to parameterize most of the PDFs. For the \mes\ distribution
of the \qqbar\ background component, we use a parameterization motivated
by phase-space arguments \cite{argus}. The following are free to vary 
in the fit: the signal and \qqbar\ background yields and charge asymmetries
\acp, along with the parameters that most strongly influence the
shape of the continuum background.

\begin{figure}[!htbp]
\begin{center}
  \includegraphics[width=1.0\linewidth]{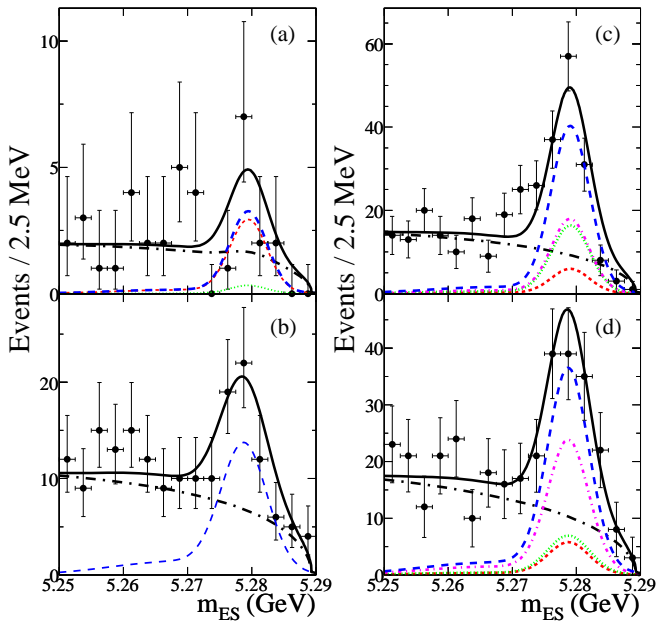}
  \caption{$B$-candidate \mes\ projections for (a) \fetaprhoz/\fetapfz, 
    (b) \fetaprhop, (c) \fetapKstz, (d) \fetapKstp.
    Color online: the solid curve is the fit function, black long-dash-dotted is the
    total background, and the blue dashed curve is the total signal
    contribution.  In (a) we separate the \rhoz\ component (red dashed)
    from the \fzero\ (green dotted). In (c,d) we separate the \Kst(892)
    (red dashed), the \KpiSwave\ (green dotted), and \Ktwost\ (magenta 
    dot-dashed) components.}
  \label{fig:proj_mes}
\end{center}
\end{figure}

We perform fits to ensembles of simulated experiments in order to 
evaluate the potential bias $Y_0$ on the fitted signal yield, which
originates from our neglect of the correlations among the variables.
Each such experiment has the same number of signal and background 
candidates as the data; given that correlations among fit variables
are negligible for \qqbar\ events, these are generated from the PDFs, while 
signal and \BB\ background events are taken from fully simulated MC 
samples. In computing the branching fraction $\cal{B}$ for each mode, we
subtract $Y_0$ from the fitted signal yield $Y$ and divide by the
selection efficiency $\varepsilon$ for signal events, the number 
of $B$ mesons in our sample, and the product of the branching 
fractions of the intermediate resonances, $\prod\calB_i$. We assume 
the branching fraction of \FourS\ to \BpBm\ and \BzBzb\ to be the 
same and equal to 50\%, consistent with the measurements \cite{PDG2008}. 
The efficiency $\varepsilon$ is evaluated from the simulation.

\begin{figure*}[!htbp]
\begin{center}
  \includegraphics[width=1.0\linewidth]{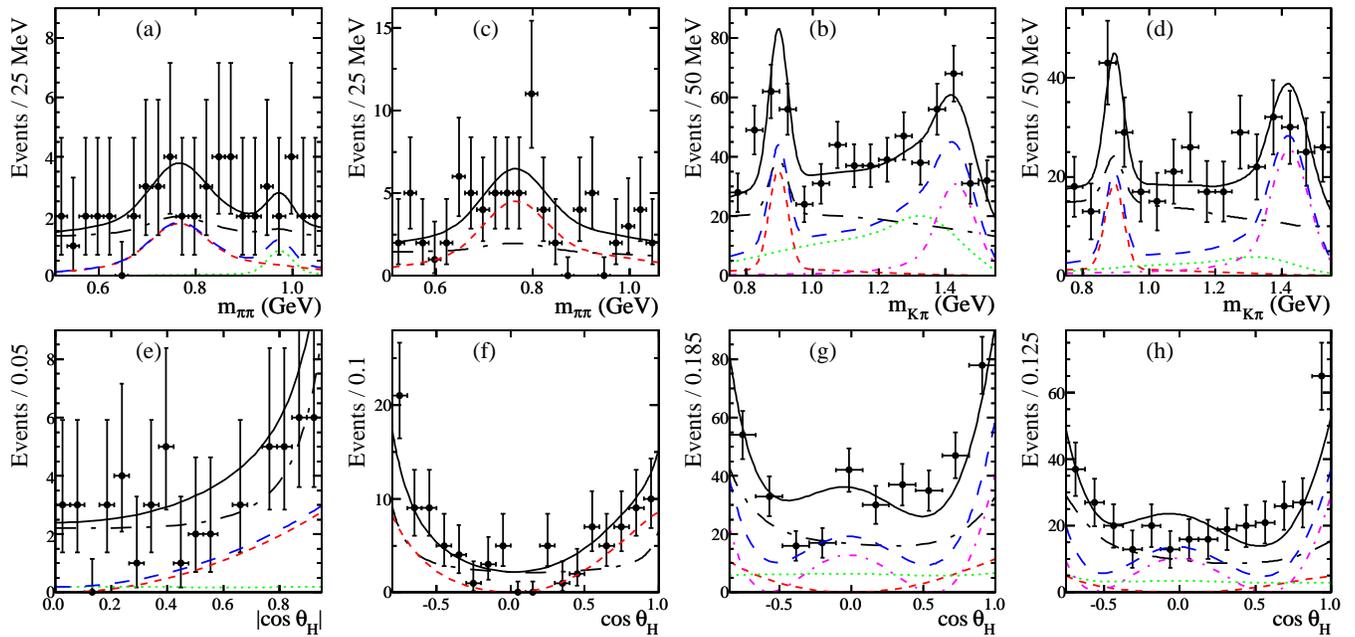}
  \caption{Top row: $B$-candidate $m_{\pi\pi}$ projections for (a) 
    \fetaprhoz/\fetapfz, (b) \fetaprhop, and $m_{K\pi}$ for (c) 
    \fetapKstz, (d) \fetapKstp; on the bottom row we plot the cosine of 
    the helicity angle $\theta_{\cal H}$ for (e) \fetaprhoz/\fetapfz, 
    (f) \fetaprhop, (g) \fetapKstz, and (h) \fetapKstp.
    Color online: the solid curve is the fit function, black 
    long-dash-dotted is the total background, and the blue dashed 
    curve is the total signal contribution.  In (a) we separate the 
    \rhoz\ component (red dashed) from the \fzero\ (green dotted). 
    In (c,d) we separate the \Kst(892) (red dashed), the \KpiSwave\ 
    (green dotted), and \Ktwost\ (magenta dot-dashed) components.}
  \label{fig:proj_mRK_hel}
\end{center}
\end{figure*}

The different submodes of \fetapKstz\ and \fetapKstp\ are combined by
adding their $-2\ln{\cal L}$ curves. For the significance of observation
$S$ we take the difference between the value of $-2\ln{\cal L}$ for the
zero signal hypothesis and the value at its minimum. For modes with
a significance below five standard deviations, we quote a 90\% 
confidence level (C.L.) upper limit, corresponding to the branching 
fraction below which lies 90\% of the total of the likelihood 
integral, in the region where the branching fraction is positive. 
The correlated and uncorrelated systematic uncertainties are 
taken into account in the above evaluations by convolving the
likelihood function given by the fitter with a Gaussian function
representing the systematic uncertainties. The results are collected
in Table \ref{tab:results}.

\begin{table*}[!bth]
\caption{
Signal yield $Y$ and its statistical uncertainty, bias $Y_0$,
detection efficiency $\epsilon$, daughter branching fraction product
$\prod\calB_i$, significance $S$ (with systematic uncertainties included), 
measured branching fraction \calB\ with statistical and systematic
errors, 90\% C.L. upper limit (U.L.), and charge asymmetry \acp.  
In the case of \fetapfz, the quoted branching fraction is the product
of \Betapfz\ with $\calB(f_0\to\pi\pi)$, which is not well known.
}
\label{tab:results}
\begin{tabular}{lccrrrccc}
\dbline
Mode            & $Y$      &$Y_0~~~~$   &$\epsilon~~~$ &$\prod\calB_i$ & $S~$       &  \calB      & \calB\ U.L. &  \acp \\
                & (events) & (events)   &(\%)$~$       & (\%)$~$       &$(\sigma)$  & $(10^{-6})$ & $(10^{-6})$ &    \\
\dbline
\fetaprhoz     & 37$\pm$15  &  9$\pm$5 & 23.4 & 17.5 & \setaprhoz\ & \retaprhoz\ & \uletaprhoz\ & $-$ \\
\sgline
\fetapfz       &  8$\pm$8   & 4$\pm$2& 25.9 & 17.5 & \setapfz\ & \retapfz\ & \uletapfz\ & $-$ \\
\sgline
\fetaprhop     & 128$\pm$22 & 15$\pm$8 & 14.3 & 17.5 & \setaprhop\ & \retaprhop\ & $-$ & \Aetaprhop\ \\ 
\sgline
\fetapKstzV     & & & & & \setapKstz\ & \retapKstz\ & \uletapKstz\ & \AetapKstz\ \\ 
~~\fetapeppKstzV  & 28$\pm$10 & 4$\pm$2 & 18.9 & 11.7 & 2.7 & $2.4^{+1.1}_{-0.9}\pm 0.3$ & & $-0.04 \pm 0.35$\\
~~\fetaprgKstzV   & 61$\pm$18 & 9$\pm$5 & 13.3 & 19.6 & 3.1 & $4.3^{+1.6}_{-1.5}\pm 0.5$ & & $\msp0.06 \pm 0.29$\\ 
\sgline
\fetapKstpV       & & & & & \setapKstp\ & \retapKstp\ & \uletapKstp\ & \AetapKstp\ \\ 
~~\fetapeppKstpVKppiz & 14$\pm$8  & 2$\pm$1 & 11.5 &  5.8 & 2.0 & $3.9^{+3.1}_{-2.1}\pm 0.5$ & & $-1.00 \pm 0.78$\\ 
~~\fetaprgKstpVKppiz  & 26$\pm$19 & 6$\pm$3 &  9.7 &  9.8 & 1.1 & $4.7^{+4.5}_{-4.1}\pm 1.3$ & & $\msp0.05 \pm 0.66$\\ 
~~\fetapeppKstpVKspip & 23$\pm$10 & 3$\pm$2 & 19.1 &  4.0 & 2.6 & $5.5^{+2.9}_{-2.4}\pm 0.7$ & & $-0.47 \pm 0.37$\\ 
~~\fetaprgKstpVKspip  & 34$\pm$15 & 10$\pm$5 & 16.2 &  6.8 & 1.6 & $4.8^{+3.2}_{-2.8}\pm 1.2$ & & $\msp0.24 \pm 0.44$\\ 
\sgline
\fetapKpiSwavez      & & & & & \setapKpiSwavez\ & \retapKpiSwavez\ & $-$ & \AetapKpiSwavez\ \\ 
~~\fetapeppKpiSwavez & 106$\pm$21 & 12$\pm$6 & 20.2 & 11.7 & 4.9 & $8.5^{+2.0}_{-1.9}\pm 1.0$ & & $-0.39 \pm 0.20$\\ 
~~\fetaprgKpiSwavez  & 115$\pm$36 & 21$\pm$11 & 17.6 & 19.6 & 2.7 & $5.8^{+2.3}_{-2.2}\pm 1.0$ & & $\msp0.32 \pm 0.31$\\ 
\sgline
\fetapKpiSwavep      & & & & & \setapKpiSwavep\ & \retapKpiSwavep\ & \uletapKpiSwavep\ & \AetapKpiSwavep\ \\ 
~~\fetapeppKpiSwavepKppiz & 36$\pm$15 & 2$\pm$1 & 13.9 &  5.8 & 2.4 & $8.8^{+4.2}_{-3.8}\pm 1.3$ & & $\msp0.00 \pm 0.41$\\ 
~~\fetaprgKpiSwavepKppiz  & 185$\pm$51 & 31$\pm$15 & 12.8 &  9.8 & 2.8 & $26.4^{+9.0}_{-8.5}\pm 5.9$ & & $\msp0.23 \pm 0.27$\\ 
~~\fetapeppKpiSwavepKspip & 18$\pm$12 & 1$\pm$1 & 18.6 &  4.0 & 1.6 & $5.1^{+3.5}_{-3.2}\pm 0.9$ & & $\msp0.13 \pm 0.59$\\ 
~~\fetaprgKpiSwavepKspip  & -29$\pm$22 & -8$\pm$4 & 17.4 &  6.8 & $-$ & $-3.8^{+4.0}_{-3.9}\pm 1.5$ & & $-0.40 \pm 1.48$\\ 
\sgline
\fetapKtstz   & & & & & \setapKtstz\ & \retapKtstz\ & $-$ & \AetapKtstz\ \\ 
~~\fetapeppKtstz & 42$\pm$13 & 2$\pm$1 & 15.1 &  5.8 & 3.7 & $9.8^{+3.4}_{-3.2}\pm 0.9$ & & $\msp0.58 \pm 0.32$\\ 
~~\fetaprgKtstz & 125$\pm$26 & 20$\pm$10 & 10.6 &  9.8 & 4.1 & $21.7^{+5.4}_{-5.3}\pm 3.0$ & & $-0.05 \pm 0.20$\\
\sgline
\fetapKtstp     & & & & & \setapKtstp\ & \retapKtstp\ & $-$ & \AetapKtstp\ \\ 
~~\fetapeppKtstpKppiz & 42$\pm$11 & 5$\pm$3 &  9.9 &  2.9 & 3.5 & $27.1^{+8.8}_{-8.1}\pm 4.5$ & & $\msp0.29 \pm 0.25$\\ 
~~\fetaprgKtstpKppiz  & 115$\pm$28 & 20$\pm$10 &  8.5 &  4.9 & 2.9 & $46.2^{+14.4}_{-13.8}\pm 12.2$ & & $-0.33 \pm 0.24$\\ 
~~\fetapeppKtstpKspip & 42$\pm$10 & 5$\pm$2 &  15.3 &  2.0 & 4.5 & $25.9^{+7.8}_{-7.1}\pm 2.7$ & & $\msp0.44 \pm 0.23$\\ 
~~\fetaprgKtstpKspip  & 62$\pm$16 & 14$\pm$7 & 12.4 &  3.4 & 3.0 & $24.1^{+8.7}_{-8.0}\pm 4.1$ & & $\msp0.22 \pm 0.25$\\ 
\dbline 
\end{tabular}
\vspace{-5mm}
\end{table*}

We show in Fig. \ref{fig:proj_mes} the data and the fit functions
projected over the variable \mes, while in Fig. \ref{fig:proj_mRK_hel}
we do the same for the $\pi\pi$ and $K\pi$ invariant masses and for
$\cos \theta_{\cal H}$. In these plots the signals are enhanced by the 
imposition of cuts on $\ln{\calL}$ and the fit variables, which retain 
40-65\% of the signal events.

We evaluate the systematic uncertainties due to the modeling of the
signal PDFs by varying the relevant parameters by their uncertainty,
derived from the consistency of fits to the above mentioned data control
samples. The fit bias arises mostly from correlations among the fit 
variables, which are modeled by the MC; the associated uncertainty is 
the sum in quadrature of half the correction itself and its statistical 
uncertainty; this is the dominant source in most cases (2.1 -- 15.4 
events), especially for the \etaptorg\ modes. We verify that the 
correlations among the variables in SXF events are the major source of 
bias by performing a dedicated study on simulated experiments in 
which the SXF component is not embedded.
The uncertainty for the SXF fraction is estimated by varying the fraction 
predicted by the MC by 2.5\% (5\%) relative for each photon (\piz) in the 
final state (most of this uncertainty originates from the simulation of 
neutral particles). We estimate the uncertainty on the charmless \BB\ 
background by repeating the fit with the yield of this component varied 
by its estimated uncertainty ($\pm$20\%).
For the $S$-wave $K\pi$ components, we vary the LASS parameters by 
the uncertainties found in \cite{LASS} (the resulting uncertainties
vary from 0.1 to 14.9 events).

Multiplicative systematic uncertainties (which do not affect the signal
significance) account for imperfect knowledge of the luminosity, tracking 
efficiency, \piz\ and \KS\ reconstruction efficiencies, and the uncertainty
on the measured branching fractions of intermediate resonances. In the
nominal fit we do not account for interference among the different
spin \Kst\ components. We estimate in a separate calculation, which
takes into account the acceptance functions of our analysis, the 
potential impact of interference when the relative strong phases
between the components are varied over the full range; we find that
the uncertainties range from 1.5\% to 14.1\%.

From the analysis of a variety of data control samples, the bias on 
\acp\ is found to be negligible for pions and --0.01 for kaons, due to 
differences of \Kp\ and \Km\ interactions in the detector
material. We correct the fitted \acp\ by +0.01 in
the modes where a charged kaon is present and assign a systematic
uncertainty of 0.02, mainly due to the bias correction.

In conclusion, we have measured the branching fraction of $B$ mesons to
nine different final states; we claim observation of four of these, 
while we obtain robust evidence for several others.
We compute the branching fraction for $\etapr K^*_0(1430)$ using
the composition of \KpiSwave\ from~\cite{Latham}. We find
\BetapKzstz=\RetapKzstz\ and \BetapKzstp=\RetapKzstp, where the
third error comes from the uncertainty in the 
$K^*_0(1430)\ra K\pi$ branching fraction \cite{PDG2008}.
No significant direct \CP-violation is seen in the investigated
channels. Our results are consistent with and supersede those reported
in \cite{babar_epKst}. Our measured branching fraction for \etaprhop\ 
favors the predictions based on pQCD and QCDF over those based 
on SCET. As in the \omegaKst\ case, we observe an enhancement of the 
tensor \Ktwost\ over the vector \Kst(892); this has not been 
anticipated by the theoretical calculations.

\par
We are grateful for the excellent luminosity and machine conditions
provided by our \pep2\ colleagues, 
and for the substantial dedicated effort from
the computing organizations that support \babar.
The collaborating institutions wish to thank 
SLAC for its support and kind hospitality. 
This work is supported by
DOE
and NSF (USA),
NSERC (Canada),
CEA and
CNRS-IN2P3
(France),
BMBF and DFG
(Germany),
INFN (Italy),
FOM (The Netherlands),
NFR (Norway),
MES (Russia),
MICIIN (Spain)
STFC (United Kingdom). 
Individuals have received support from the
Marie Curie EIF (European Union),
the A.~P.~Sloan Foundation (USA),
the Binational Science Foundation (USA-Israel), and the University of
Colorado at Boulder Graduate School Faculty Fellowship program.


\renewcommand{\baselinestretch}{1}

\end{document}